\documentclass[useAMS,usenatbib,usegraphicx]{mn2e}
\usepackage{ulem,color}

\title[]{Modelling the 3D morphology and proper motions of the planetary nebula NGC 6302}
\author[]{ L. Uscanga,$^{1}$\thanks{E-mail:lucero@astro.noa.gr}
  P. F. Vel\'azquez,$^{2}$ A. Esquivel,$^{2}$  A. C. Raga,$^{2}$ 
P. Boumis,$^{1}$ 
\newauthor
 and  J. Cant\'o,$^{3}$\\ 
$^{1}$Institute of Astronomy, Astrophysics, Space Applications and Remote Sensing, National Observatory of Athens, \\
15236 Athens, Greece\\
$^{2}$Instituto de Ciencias Nucleares, UNAM, Apartado Postal 70-543, 04510 M\'exico, DF, Mexico\\
$^{3}$Instituto de Astronom\'{\i}a, UNAM, Apartado Postal 70-264, 04510 M\'exico, DF, Mexico}

\begin{document}


\pagerange{\pageref{firstpage}--\pageref{lastpage}} \pubyear{2014}

\maketitle

\label{firstpage}

\begin{abstract}
We present 3D hydrodynamical simulations of an isotropic fast wind
interacting with a previously ejected toroidally-shaped slow wind in
order to model both the observed morphology and the kinematics of the
planetary nebula (PN) \mbox{NGC 6302}. This source, also known as the
Butterfly nebula, presents one of the most complex morphologies ever
observed in PNe.  From our numerical simulations, we have obtained an
intensity map for the H$\alpha$ emission to make a comparison with the
\textit{Hubble Space Telescope} (\textit{HST}) observations of this
object.  We have also carried out a proper motion (PM) study from our
numerical results, in order to compare with previous observational
studies.
We have found that the two interacting stellar wind model 
reproduces well the morphology of \mbox{NGC 6302}, and while the PM
in the models are similar to the observations, our results suggest that
an acceleration mechanism is needed to explain the Hubble-type expansion
found in \textit{HST} observations.

\end{abstract}


\section{Introduction}

Planetary nebulae (PNe) form around stars of low and intermediate mass
($\le 8$~M$_{\sun}$) towards the end of their evolution.  The nebula
originates from the interaction between a fairly dense ($\sim$10$^{-5}$
M$_{\sun}$~yr$^{-1}$) `slow wind' ($\sim$10 km s$^{-1}$) lost by the
star as a red giant on the asymptotic giant branch (AGB) with a
tenuous ($\sim$10$^{-8}$ M$_{\sun}$~yr$^{-1}$) `fast wind'
($\sim$1000~km~s$^{-1}$) that follows after the AGB phase. The
nebula material becomes ionized once the effective temperature of the
central star, which evolves into a white dwarf, exceeds 25000 K (for a
review see \citealt{bal02}).

NGC 6302 (PN G349.5$+$01.0) is a bipolar PN. 
Its complex and clumpy morphology
shown in H$\alpha$ and [N\,{\sc ii}] \textit{Hubble Space Telescope}
(\textit{HST}) images \citep{szy11} can be roughly approximated by a
bipolar shape with the two main lobes extending in the east-west
direction. It presents a highly pinched waist which is characteristic
for butterfly-shaped bipolar PNe \citep{bal02}.  The central star of
NGC 6302 was directly detected for the first time by \citet{szy09}. It
was partially obscured by a dense equatorial lane \citep{mat05} and
molecular material observed in CO emission \citep{per07, din08} that
is tracing an expanding torus oriented in the north-south direction,
approximately perpendicular to the optical nebula axis.  The distance
to this nebula was not known accurately. \citet{gom89} gave a firm
lower limit of 0.8$\pm$0.3~kpc based on a radio expansion
proper-motion (PM) measurement of the nebular core, and estimate a
distance of 2.2$\pm$1.1~kpc from measurements of pressure-broadening
of radio recombination lines.  Lately, \citet{mea08} determined an
unambiguous distance of 1.17$\pm$0.14~kpc, derived from expansion PMs
of 15 knots located at the northwestern lobe of the nebula. These PMs
were measured comparing two ground-based images with a time separation
of $\sim$51~yr, indicating a Hubble-type expansion. Hubble-type
outflows are characterised by an expansion velocity that increases
linearly with the distance to the central star. This type of outflow
has been observed also in other objects, i.e., the nebula around the
symbiotic system Hen 2-104 \citep{cor01}, NGC 6537 \citep{cor93}, Mz 3
\citep{red00}, see \citet{cor04} for other examples.

Recently, Szyszka et al. (2011) measured the expansion PMs in NGC 6302,
comparing two \textit{HST} images in [N\,{\sc ii}] separated by 9.4 yr. The
velocity field follows a Hubble law in agreement with the previous
results of \citet{mea08}. The PM vectors present a pattern mostly
radial pointing back to the central source with a position close to
the central star detected by \citet{szy09}. Their
results show that the lobes of NGC 6302 were ejected during a brief
event 2250$\pm$35 yr ago (in agreement with the result of
\citealt{mea08}), and they find evidence for a subsequent 
acceleration that increased the velocity of the inner regions 
by 9.2~km~s$^{-1}$, possibly related to the onset of the ionization.

In this work, we investigate whether or not the morphological and
kinematical characteristics of NGC 6302 can result from the
interaction between an isotropic fast wind with a toroidally-shaped
slow wind.  We have also considered a clumpy structure for the slow
wind, based on observations of AGB shells, which look like clumpy and
filamentary shell structures \citep{cox12,ste13}.  The paper is
organised as follows. In Section 2, we describe the basic assumptions
and the initial conditions used in the numerical simulations. In
Section 3, we present the results including synthetic H$\alpha$
emission maps, simulated PMs of nebular knots, and their comparison
with the observations. Finally, in Section 4, we discuss the
implications of our results and give our conclusions.

\section[]{Initial conditions of the numerical simulations}
Our simulations are based on the generalisation of the interacting
stellar winds model (GISW) in which, an isotropic fast wind launched by
a star expands into a previously ejected toroidally-shaped slow wind
\citep{ick89,mel95b}. We assumed that
the slow wind has a density distribution with a high contrast between
the equator and the pole, which is described by
the following equation (given by \citealt{mel95a})
\begin{equation}
\rho(r,\theta)=\rho_0 g(\theta)\bigg(r_0/r\bigg)^2,
\label{rhowind}
\end{equation}
with
\begin{equation}
g(\theta)=1-\alpha\bigg[ \frac{1-\mathrm{exp}(-2\beta
  \mathrm{cos}^2\theta)}{1-\mathrm{exp}(-2\beta)}\bigg],
\end{equation}
where $r$ is the distance from the central star and $\theta$ is the
polar angle  ($\theta=0\degr$ at the pole, and $90\degr$ at
the equator).  The parameter $\alpha$ determines the ratio between the value of the
density at the equator and that at the pole, while $\beta$ determines the
shape of the variation, and therefore that of the slow wind
(see \citealt{mel95a}).  The value of $\rho_0$ can be calculated from the
mass-loss rate $\dot{M}_{AGB}$ as:
\begin{equation}
 \rho_0=\dot{M}_{AGB}/(4\pi\ r^2_0 v_{AGB}),
\end{equation}
where $v_{AGB}$ is the constant terminal velocity of the AGB wind and
$r_0$ is the radius of the region where the stellar wind is imposed.

The 3D numerical simulations were performed with the {\sc{yguaz\'u}}
hydrodynamical code \citep{rag00}, which integrates the gas dynamical
equations with a second-order accurate scheme (in time and space)
using the `flux-vector splitting' method of \citet{van82} on a binary
adaptive grid. A rate equation for neutral hydrogen is integrated
together with the gas dynamics equations to include the radiative
losses through a parametrized cooling function that depends on the
density, temperature and hydrogen ionization fraction \citep{rag04}.

To model the NGC 6302 nebula, we have used a computational domain of
$(1.5, 1.5, 3.0)\times 10^{18}$~cm along the $x$-, $y$-, and $z$-axes,
respectively. Five refinement levels were allowed in the adaptive
Cartesian grid, achieving a resolution of $5.9\times 10^{15}$~cm at
the finest level.  
For the initial condition we have filled the computational domain with
the density distribution obtained from equation (1)-(3) with 
$\dot{M}_{AGB}$, $v_{AGB}$,  and $r_0$ equal to $5\times
10^{-4}$~M$_{\sun}$~yr$^{-1}$, $15~{\mathrm{km\ s^{-1}}}$, and 
$3.8\times 10^{16}$~cm, respectively.  We have used $\alpha=0.999$ and
$\beta=10$ in order to reproduce the observed morphology.

To simulate a `clumpy' circumstellar medium (CSM), we have
  modulated the density by a fractal structure with a spectral index of
$-11/3$, which is consistent with a turbulent interstellar medium
(see \citealt{esq05,esq03,oss06}).  This clumpy density is imposed on 
the  initial
  condition and has fluctuations on the order of $20\%$ of the mean
density value.  In this CSM, we start an isotropic 
and fast stellar wind with mass injection $\dot{M}_f$ of
$10^{-7}$~M$_{\sun}$~yr$^{-1}$, and a velocity $v_f$ of
$1800~{\mathrm{km\ s^{-1}}}$. 

\section[]{Results}

\subsection{Synthetic H$\alpha$ emission maps}

The 3D hydrodynamical simulations of two interacting winds were carried out
with the setup given in Section 2. 
The fast wind generates a shock wave which propagates into the
surrounding CSM. A global bipolar morphology is generated and a
similar size to the observed one is achieved after a time
of 2000~yr, which is in agreement with the dynamical age given by
 Szyszka et al. (2011).

\begin{figure}
\includegraphics[width=82mm]{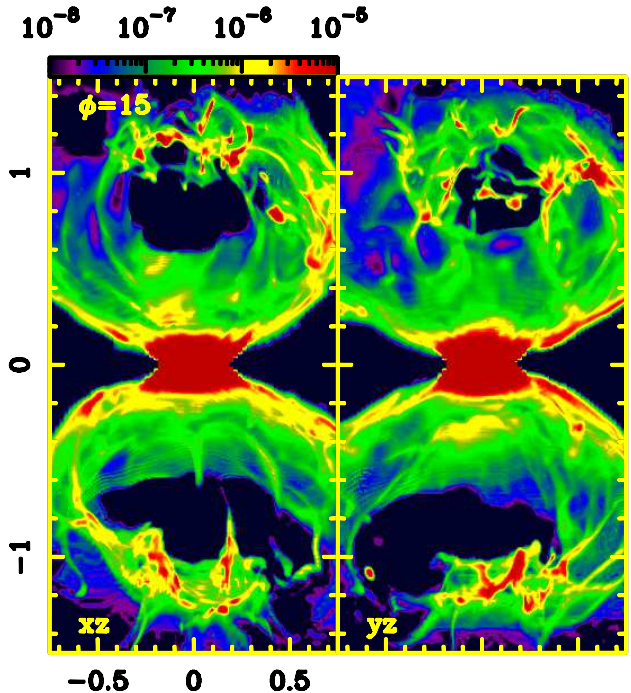}
 \caption{Synthetic H$\alpha$ emission maps obtained at an integration
   time of 2000 yr and for the $xz-$ and $yz-$ projections (left and
   right panel, respectively).  The angle $\phi$ is the angle between
   the nebula axis (the $z-$ axis of the computational domain) and the plane of the sky.  The vertical and
   horizontal axes are in units of 10$^{18}$ cm. The horizontal colour
   bar gives the H$\alpha$ flux in units of
   erg~s$^{-1}$~cm$^{-2}$~sr$^{-1}$.}
\label{fha}
\end{figure}

From the density and temperature distribution given by our 3D
hydrodynamical simulations, we can perform synthetic H$\alpha$
maps. In Fig. \ref{fha}, we show the synthetic H$\alpha$ emission maps
obtained for the $xz-$ and $yz-$ projections (left and right panels of
Fig. \ref{fha}, respectively), using an angle $\phi=15\degr$. This
angle $\phi$ is the angle between the nebula axis (the $z$ axis of the
computational domain) and the plane of the sky, in agreement with 
previous observational results \citep{per07,mea08,din08}. 
Fig. \ref{fha} shows a bipolar morphology for both
projections, with a waist of $6\times 10^{17}$~cm, which has a similar
size to the observed one ($31\arcsec$ if a distance of 1.17 kpc is
considered).  Furthermore, both panels display a filamentary and
clumpy structure, which is a consequence of the interaction between
the fast wind and the clumpy slow wind.

\subsection{Simulated PMs}
To calculate the PMs of the nebular knots we have re-started our
simulation at an integration time of 2000 yr, and leave it to evolve
by 10 additional yr.  Then, we obtain the PMs of the knots found in
the simulation using a cross-correlation method. We have used the ``PM
mapping'' technique described by \citet{rag13}, in which, the PMs from
pair of images are derived by defining boxes including emitting knots
and carrying out a cross-correlation function of the emission within
the boxes. The PM is  obtained from a fit
to the peak of the cross-correlation function. This method has proven
to be better than carrying out direct fits (e.g., a Gaussian or a
two-dimensional paraboloid) to the observed emission features, because
the cross-correlation functions, which are integrals of the emission
within the chosen boxes, have higher signal-to-noise ratios than the
images.

The cross-correlation boxes have a size of 30$\times$30 pixels
equivalent to 10$\times$10 arcsec$^2$. For a given box of size $L$, we
first check whether or not the condition $f\ge f_{min}$ is satisfied
in at least one pixel within the central inner box of size $L/2$ (see
appendix of \citealt{rag13}). Here $f$ is the H$\alpha$ flux of the
image and $f_{min}$ is set equal to $10^{-7}$
erg~s$^{-1}$~cm$^{-2}$~sr$^{-1}$. If this condition is met at least
for a single pixel in each of the two epochs that are being analysed,
the cross-correlation function (within the $L$-size box in the two
images) and the PM are computed.

\begin{figure}
\includegraphics[width=82mm]{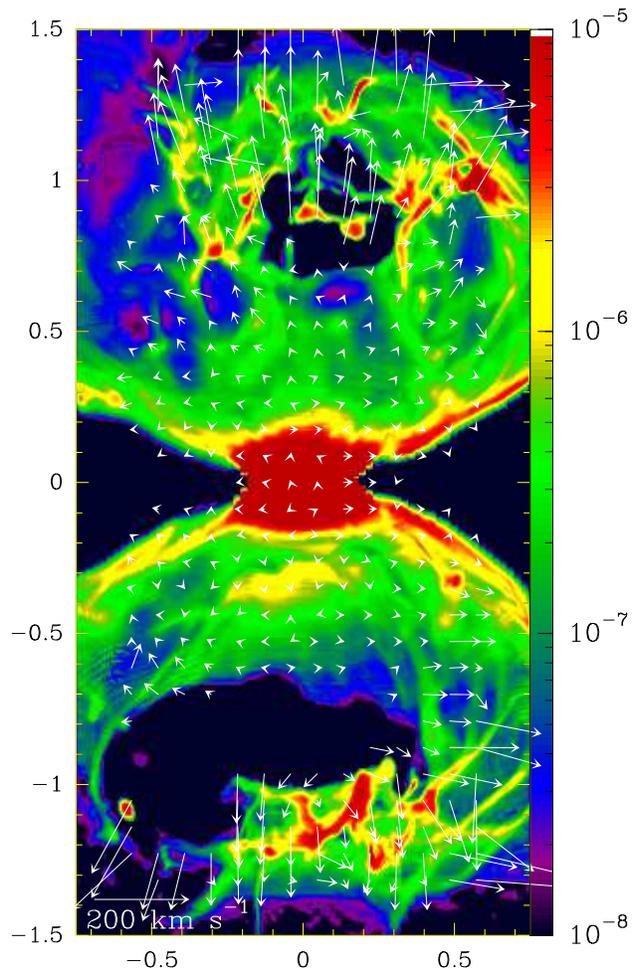}
 \caption{Relative H$\alpha$ PM map of NGC 6302.  The nebula axis is
   oriented at $15\degr$ with respect to the plane of the sky.  The
   vertical and horizontal axes have the same dimensions as in
   Fig. \ref{fha}.  The vertical colour bar gives the H$\alpha$ flux in
   units of erg~s$^{-1}$~cm$^{-2}$~sr$^{-1}$.  The length of each
   arrow indicates the relative PM of the H$\alpha$ emission knots
   computed in 10$\times$10 arcsec$^2$ boxes. The arrow at bottom left
   corresponds to a value of 200~km~s$^{-1}$.}
\label{fhapm}
\end{figure}

Fig. \ref{fhapm} shows the $yz-$ projection of the synthetic H$\alpha$ map
overlaid by white arrows which represent PM vectors calculated with
the ``PM mapping'' technique.
Interestingly, the PMs show deviations from 
the radial direction mostly for filamentary features in the outer regions 
of the outflow lobes. For a spatially extended filamentary structure, 
the PM perpendicular to the locus of the emission is well determined. 
While the component of the PM of the flow along the 
filament is highly uncertain, resulting in a large dispersion of the PM
vectors determined for filamentary features.

\begin{figure*}
\includegraphics[width=164mm]{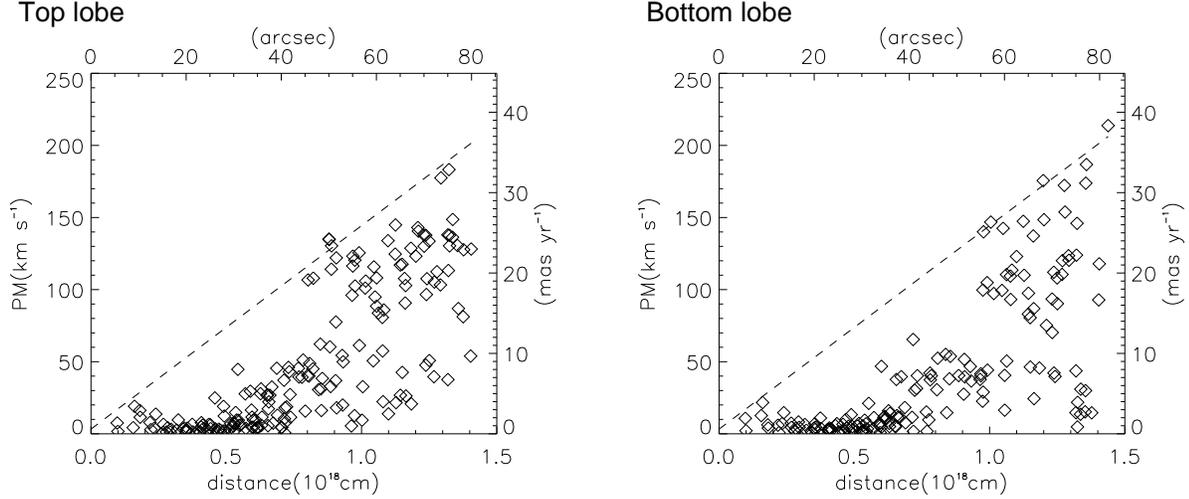}
 \caption{The PMs shown in Fig. \ref{fhapm} are plotted versus
   separation from the central source.  The nebula axis is oriented at
   $15\degr$ with respect to the plane of the sky.  The left panel
   displays the PMs for $z>0$ (the top lobe of the synthetic nebula),
   while the right panel shows the PMs for $z<0$ (bottom lobe of the
   synthetic nebula). The diagonal line represents
   the linear fit of the observed PMs done by Szyszka et al. (2011). }
\label{figpms}
\end{figure*}

\subsection{Comparison between the simulated PMs and the observed PMs}

Fig. \ref{figpms} shows PMs vs. distance to the central star plots
for the top lobe (left panel) and bottom lobe (right panel). From this
figure, we note that the more distant filaments and knots move faster
than those located near the source.  However, the PMs
obtained from our model do not resemble the behaviour given by a
Hubble-type expansion, such as found in the NGC 6302 observations
(Szyszka et al. 2011). The Hubble-type expansion is represented in both panels
of Fig. \ref{figpms} by dashed lines. We can see
  from the Fig. \ref{figpms} that many of the knots far from the
  source have velocities comparable to the observations, however we see a
  majority with velocities below the Hubble-type expansion.  At the
  same time, the knots that are closer to the source ($\leq 5\times
  10^{17}~\mathrm{cm}$), have a remarkably constant PM, well below the
  observations.  This result suggest that an additional acceleration
  mechanism is needed, particularly near the central star, which is
  consistent with the findings of Szyszka et al. (2011).  A possible candidate
  for this acceleration mechanism is the ionizing flux from the
  central star, which would photoevaporate the clumps and push them
  through a rocket effect \citep[see][]{esq07}. This, and any other
  acceleration mechanism is not included in our models, and will be
  pursued in a following work.

\section{Discussion and Conclusions}

We have carried out 3D hydrodynamical simulations of a two
interacting-winds model, in order to explain both the global
morphology and the tangential velocity of the PN NGC 6302. 
We reproduce the morphology and size of the nebula after an
integration time of 2000 yr, which is similar to the dynamical time
estimate given by Szyszka et al. (2011).  
We note the that it was necessary to use a high equator
to pole density contrast of $\sim$1000 to obtain the observed
morphology.

A PM study was done on the synthetic H$\alpha$ images during a time
span of 10 yr similar to the time span between observed images. The
magnitude of the PMs as well as the separation from the central star
are quantitatively similar, although the obtained PMs do not follow a
Hubble-type expansion as the observations indicate.
This suggests that an additional acceleration
  mechanism is acting, in  particularly near the central star where
  the PM discrepancy is larger.
  A possible candidate for this mechanism is the rocket effect from
the photoevaporation of the clumps, due to the radiation of the
central star.

\section*{Acknowledgments}
We would like to dedicate this paper to the memory of our dear
colleague Yolanda G\'omez, who sadly passed away in 2012. She was also
involved in this project. LU and PB are grateful to John Meaburn
for his comments related to a very early draft version of this
paper. 
We are thankful to our referee, Myfanwy Lloyd, for her valuable 
comments on the manuscript.
LU acknowledges support from grant PE9-1160 of the Greek
General Secretariat for Research and Technology in the framework of
the program Support of Postdoctoral Researchers. PFV, ACR, JC, and AE
acknowledge support from the CONACyT grants 
167611, 167625, and UNAM DGAPA grant IG100214.

\bsp

\label{lastpage}


\begin{thebibliography}{99}
\bibitem[\protect\citeauthoryear{Balick \& Frank}{2002}]{bal02} Balick B., Frank A., 2002, ARA\&A, 40, 439 
\bibitem[\protect\citeauthoryear{Corradi}{2004}]{cor04} 
Corradi R.~L.~M., 2004, ASPC, 313, 148 
\bibitem[\protect\citeauthoryear{Corradi et 
al.}{2001}]{cor01} Corradi R.~L.~M., Livio M., Balick B., 
Munari U., Schwarz H.~E., 2001, ApJ, 553, 211  
\bibitem[\protect\citeauthoryear{Corradi \& Schwarz}{1993}]{cor93} Corradi R.~L.~M., Schwarz H.~E., 1993, A\&A, 269, 462
\bibitem[\protect\citeauthoryear{Cox et al.}{2012}]{cox12} Cox N.~L.~J., et al., 2012, A\&A, 537, A35 
\bibitem[\protect\citeauthoryear{Dinh-V-Trung et al.}{2008}]{din08} Dinh-V-Trung, Bujarrabal V., Castro-Carrizo A., Lim J., Kwok S., 2008, ApJ, 673, 934 
\bibitem[\protect\citeauthoryear{Esquivel \& Lazarian}{2005}]{esq05} Esquivel A., Lazarian A., 2005, ApJ, 631, 320 
\bibitem[\protect\citeauthoryear{Esquivel et al.}{2003}]{esq03} Esquivel A., Lazarian A., Pogosyan D., Cho J., 2003, MNRAS, 342, 325 
\bibitem[\protect\citeauthoryear{Esquivel \& Raga}{2007}]{esq07} Esquivel A., Raga A.~C., 2007, MNRAS, 377, 383 
\bibitem[\protect\citeauthoryear{G\'omez et al.}{1989}]{gom89} G\'omez Y., Rodr\'iguez L.~F., Moran J.~M., Garay G., 1989, ApJ, 345, 862 
\bibitem[\protect\citeauthoryear{Icke, Preston, \& Balick}{1989}]{ick89} Icke V., Preston H.~L., Balick B., 1989, AJ, 97, 462 
\bibitem[\protect\citeauthoryear{Matsuura et al.}{2005}]{mat05} Matsuura M., Zijlstra A.~A., Molster F.~J., Waters L.~B.~F.~M., Nomura H., Sahai R., Hoare M.~G., 2005, MNRAS, 359, 383 
\bibitem[\protect\citeauthoryear{Meaburn et al.}{2008}]{mea08} Meaburn J., Lloyd M., Vaytet N.~M.~H., L{\'o}pez J.~A., 2008, MNRAS, 385, 269 
\bibitem[\protect\citeauthoryear{Mellema}{1995}]{mel95a} Mellema G., 1995, MNRAS, 277, 173 
\bibitem[\protect\citeauthoryear{Mellema \& Frank}{1995}]{mel95b} Mellema G., Frank A., 1995, MNRAS, 273, 401 
\bibitem[\protect\citeauthoryear{Ossenkopf et al.}{2006}]{oss06} Ossenkopf V., Esquivel A., Lazarian A., Stutzki J., 2006, A\&A, 452, 223 
\bibitem[\protect\citeauthoryear{Peretto et al.}{2007}]{per07} Peretto N., Fuller G., Zijlstra A., Patel N., 2007, A\&A, 473, 207 
\bibitem[\protect\citeauthoryear{Raga, Navarro-Gonz{\'a}lez, \& Villagr{\'a}n-Muniz}{2000}]{rag00} Raga A.~C., Navarro-Gonz{\'a}lez R., Villagr{\'a}n-Muniz M., 2000, RMxAA, 36, 67 
\bibitem[\protect\citeauthoryear{Raga et al.}{2013}]{rag13} 
Raga A.~C., Noriega-Crespo A., Carey S.~J., Arce H.~G., 2013, AJ, 145, 28 
\bibitem[\protect\citeauthoryear{Raga \& Reipurth}{2004}]{rag04} Raga A.~C., Reipurth B., 2004, RMxAA, 40, 15 
\bibitem[\protect\citeauthoryear{Redman et al.}{2000}]{red00} 
Redman M.~P., O'Connor J.~A., Holloway A.~J., Bryce M., Meaburn J., 2000, 
MNRAS, 312, L23 
\bibitem[\protect\citeauthoryear{Steffen et al.}{2013}]{ste13} Steffen W., Koning N., Esquivel A., Garc{\'{\i}}a-Segura G., Garc{\'{\i}}a-D{\'{\i}}az M.~T., L{\'o}pez J.~A., Magnor M., 2013, MNRAS, 436, 470  
\bibitem[\protect\citeauthoryear{Szyszka et al.}{2009}]{szy09} Szyszka C., Walsh J.~R., Zijlstra A.~A., Tsamis Y.~G., 2009, ApJ, 707, L32 
\bibitem[\protect\citeauthoryear{Szyszka, Zijlstra, \& Walsh}{2011}]{szy11} Szyszka C., Zijlstra A.~A., Walsh J.~R., 2011, MNRAS, 416, 715 
\bibitem[\protect\citeauthoryear{van Leer}{1982}]{van82} van Leer B., 1982, Lecture Notes in Physics, 170, 507 

\end{thebibliography}
\end{document}